\documentstyle[aps,preprint]{revtex}
\begin{document}
\draft


\preprint{YITP-98-61, gr-qc/9809050}

\title{On the Noether charge form of the first law of black hole mechanics} 
\author{Shinji Mukohyama}
\address{
Yukawa Institute for Theoretical Physics, 
Kyoto University \\
Kyoto 606-8502, Japan
}
\date{\today}

\maketitle


\begin{abstract} 

The first law of black hole mechanics was derived by Wald in a
general covariant theory of gravity for stationary variations around a
stationary black hole. It is formulated in terms of Noether charges, 
and has many advantages. 
In this paper several issues are discussed to strengthen the validity 
of the Noether charge form of the first law.
In particular, a gauge condition used in the derivation is 
justified.
After that, we justify the generalization to non-stationary 
variations done by Iyer-Wald. 

\end{abstract}


\section{Introduction}
	\label{sec:intro}

Analogy with thermodynamics is one of the most interesting results in
the theory of black holes. It can be summarized as laws of black hole 
mechanics~\cite{BCH1973}. In particular, the first law makes it
possible to assign entropy to horizon area and temperature to surface
gravity up to a coefficient~\cite{Bekenstein1973}. It was Hawking's 
discovery~\cite{Hawking1975} of thermal radiation from a black hole
that determined the coefficient.

In Ref.~\cite{Wald1993}, the first law of black hole mechanics was
derived in a general covariant theory of gravity for stationary
variations around a stationary black hole. It is formulated as a
relation among variations of those quantities such as energy, angular
momentum and entropy, each of which is defined in terms of a Noether
charge.
The first law was extended to non-stationary variations around a
stationary black hole in Ref.~\cite{Iyer&Wald1994}.

These first laws in the Noether charge form have many advantages over
the original first law.
For example, it gives a general method to calculate stationary black
hole entropy in general covariant theories of
gravity~\cite{Iyer&Wald1994};
it connects various Euclidean methods for computing black 
hole entropy~\cite{Iyer&Wald1995}; 
it suggests a possibility of defining  entropy of non-stationary black 
holes~\cite{Iyer&Wald1994,JKM1994};  etc.

However, in their derivation there are several issues to 
be discussed in more detail. 
\begin{enumerate}
 \renewcommand{\labelenumi}{(\alph{enumi})}
 \item	In Ref.~\cite{Wald1993}, unperturbed and perturbed stationary 
	black holes are identified so that horizon generator Killing 
	fields with unit surface gravity coincide near the horizons 
	and that stationary Killing fields and axial Killing fields 
	coincide at infinity. This corresponds to taking a certain 
	gauge condition in linear perturbation theory. 
	For a complete understanding of the first law, we have to 
	clarify whether such a gauge condition can be 
	imposed or not. If it can, then we like to know whether such
	a gauge condition is necessary. 
	Note that, on the contrary, the original derivation in general 
	relativity by Bardeen, Carter and Hawking~\cite{BCH1973} is 
	based on a gauge condition such that the stationary Killing 
	fields and the axial Killing fields coincide everywhere on a 
	spacelike hypersurface whose boundary is a union of a horizon 
	cross section and spatial infinity.
 \item	In Ref.~\cite{Iyer&Wald1994}, the first law is extended to 
	non-stationary perturbations around a stationary black hole. 
	In the derivation, change of black hole entropy is calculated
	on a ($n-2$)-surface, which is a bifurcation surface for 
	an unperturbed black hole, but which is not a cross section 
	of an event (nor apparent) horizon for a perturbed 
	non-stationary black hole in general. 
	Does this mean that black hole entropy would be assigned to a 
	surface which is not a horizon cross section for a 
	non-stationary black hole? 
	It seems more natural to assign black hole entropy to 
	a horizon cross section also for a non-stationary black hole. 
\end{enumerate}
In this paper these two issues are discussed and it is concluded
that there are no difficulties in the derivation of the Noether charge 
form of the first law for both stationary and non-stationary 
perturbations about a stationary black hole. 
On the way, we give an alternative derivation of the first law based 
on a variation in which a horizon generator Killing field with unit 
surface gravity is fixed.

In Sec.~\ref{sec:gauge} gauge conditions near horizon are investigated. 
In Sec.~\ref{sec:stationary} the first law of black holes is derived 
for stationary variations around a stationary black hole. 
In Sec.~\ref{sec:non-stationary} the derivation is extended to 
non-stationary variations around a stationary black hole. 
Sec.~\ref{sec:summary} is devoted to a summary of this paper.

\section{Gauge conditions}
	\label{sec:gauge}

Consider a stationary black hole in $n$-dimensions, which has a
bifurcating Killing horizon. 
Let $\xi^a$ denote a generator Killing field of the Killing horizon, which 
is normalized as $\xi^a=t^a+\Omega_H^{(\mu)}\varphi_{(\mu)}^a$, 
and $\Sigma$ be the bifurcation surface. Here, $t^a$ is the Killing 
field of stationarity with unit norm at infinity, 
$\{\varphi_{(\mu)}^a\}$ ($\mu=1,2,\cdots$) is a family of axial Killing 
fields, and $\{\Omega_H^{(\mu)}\}$ is a family of constants (angular 
velocities).

Now let us show that it is not possible in general to impose a gauge 
condition such that $\delta\xi^a=0$ near the bifurcation surface. 
For this purpose we shall temporarily assume that $\delta\xi^a=0$ and 
show a contradiction.

On $\Sigma$, the covariant derivative of $\xi^a$ is given by
%
\begin{equation}
 \nabla_b\xi^a = \kappa{\bf\epsilon}_b^{\ a},
\end{equation}
where $\kappa$ is the surface gravity corresponding to $\xi^a$ and
${\bf\epsilon}_{ab}$ is binormal to $\Sigma$. 
However, the variation of the l.h.s. is zero:
%
\begin{equation}
 \delta(\nabla_b\xi^a) = \delta\Gamma^a_{bc}\xi^c =0
 	\label{eqn:delta-d-xi-0}
\end{equation}
since $\xi^a=0$, where $\delta\Gamma^a_{bc}$ is given by
%
\begin{equation}
 \delta\Gamma^a_{bc} = 
	\frac{1}{2}g^{ad}(\nabla_c\delta g_{db}
	+\nabla_b\delta g_{dc}-\nabla_d\delta g_{bc}).
\end{equation}
Hence, 
%
\begin{equation}
 \delta{\bf\epsilon}_b^{\ a} = 
 	-\frac{\delta\kappa}{\kappa}{\bf\epsilon}_b^{\ a}.
\end{equation}
Substituting this into the identity 
$\delta({\bf\epsilon}_b^{\ a}{\bf\epsilon}^b_{\ a})=0$, we obtain
%
\begin{equation}
 0 = \delta({\bf\epsilon}_b^{\ a}{\bf\epsilon}^b_{\ a}) = 
	-\frac{4\delta\kappa}{\kappa}.\label{eqn:delta-kappa-0}
\end{equation}

Thus, the assumption $\delta\xi^a=0$ leads $\delta\kappa=0$, 
which implies, for example, that $\delta M=0$ for the vacuum general 
relativity in staticity, where $M$ is mass of Schwarzschild black holes. 
This peculiar behavior can be understood as appearance of a coordinate 
singularity at the bifurcation surface of a coordinate fixed by the
gauge condition $\delta\xi^a=0$ since in the above argument finiteness
of $\delta\Gamma^a_{bc}$ has been assumed implicitly. 
Therefore, it is impossible to impose the condition $\delta\xi^a=0$
near the bifurcation surface whenever $\delta\kappa\ne 0$.

As mentioned above, the original derivation of the first law in 
Ref.~\cite{BCH1973} adopt the gauge condition 
$\delta t^a=\delta\varphi^a=0$. This leads $\delta\xi^a=0$ when 
$\delta\Omega=0$ (for example, when we consider static black holes).
Of course, in Ref.~\cite{BCH1973}, a general horizon cross section 
(not necessary a bifurcation surface) is considered as a surface on 
which black hole entropy is calculated. 
Hence, the above argument arises no difficulties unless the cross 
section is taken to be the bifurcation surface. 
The derivation in Ref.~\cite{BCH1973} suffers from the above argument
if and only if black hole entropy is estimated on the bifurcation 
surface.

On the other hand, arguments like the above do not lead to any 
contradiction if we adopt a gauge condition such that $\tilde{\xi}^a$ 
is fixed near the bifurcation surface under variations, 
where $\tilde{\xi}^a=\xi^a/\kappa$ is a horizon generator Killing field
with unit surface gravity. 
Moreover, it is concluded that, if we intend
to fix a horizon generator Killing field, then it must have the same 
value of surface gravity for unperturbed and perturbed black holes. 
Hence, the gauge condition $\delta\tilde{\xi}^a=0$ near the bifurcation 
surface adopted in Ref.~\cite{Wald1993,Iyer&Wald1994} is very natural 
one.

In fact, it is always possible to identify unperturbed 
and perturbed stationary black holes so that the Killing horizons
and the generator Killing fields with unit surface gravity
coincide. 
As stated in Ref.~\cite{Wald1993}, such an identification can be done 
at least in a neighborhood of the horizon by using the general formula 
for Kruskal-type coordinates $(U,V)$ given in Ref.~\cite{Racz&Wald1992}. 
(The identified Killing horizon is given by $U=0$ and $V=0$. The
identified Killing field with unit surface gravity is given by 
$\tilde{\xi}^a=U(\partial/\partial U)^a-V(\partial/\partial V)^a$.)

The purpose of the next section is to discuss about the remaining
gauge condition $\delta t^a=\delta\varphi^a=0$ at infinity. 
It is evident that this gauge condition at infinity can be imposed by
identifying the perturbed and unperturbed specetimes suitably. 
So, our question now is whether this gauge condition is necessary or
not. 
For this purpose we temporarily adopt a gauge condition such that
$\tilde{\xi}^a$ is fixed everywhere on a hypersurface connecting the
bifurcation surface and spatial infinity. 
In deriving the first law in this gauge condition, the gauge condition
$\delta t^a=\delta\varphi^a=0$ at infinity is found to be 
necessary for a proper interpretation of the first law. 
On the other hand, as shown in Sec.~\ref{sec:non-stationary},
it is not necessary to fix $\tilde{\xi}^a$ near the bifurcation surface, 
strictly speaking. 
Hence, it can be concluded that the minimal set of gauge conditions
necessary for the derivation of the first law is that $t^a$ and
$\varphi^a$ are fixed at spatial infinity.

\section{The first law for stationary black holes}
	\label{sec:stationary}
	
Before deriving the first law, we review basic ingredients of the 
formalism.

In this paper, we consider a classical theory of gravity in $n$ dimensions 
with arbitrary matter fields, which is described by a diffeomorphism 
invariant Lagrangian $n$-form ${\bf L}(\phi)$, where $\phi$ denotes 
dynamical fields in the sense of Ref.~\cite{Iyer&Wald1994}.


The Noether current ($n-1$)-form ${\bf j}[V]$ for a vector field
$V^a$ is defined by 
%
\begin{equation}
 {\bf j}[V] \equiv 
 	{\bf\Theta}(\phi,{\cal L}_V\phi)-V\cdot{\bf L}(\phi),
 	\label{eqn:j-def}
\end{equation}
where the ($n-1$)-form ${\bf\Theta}(\phi,\delta\phi)$ is defined by 
%
\begin{equation}
 \delta{\bf L}(\phi) = 
 	{\bf E}(\phi)\delta\phi + d{\bf\Theta(\phi,\delta\phi)}.
 	\label{eqn:Theta-def}
\end{equation}
It is easily shown that the Noether current is closed as 
%
\begin{equation}
 d{\bf j}[V] = -{\bf E}(\phi){\cal L}_V\phi = 0,
\end{equation}
where we have used the equations of motion ${\bf E}(\phi)=0$.
Hence, by using the machinery developed in Ref.~\cite{Wald1990}, we 
obtain the Noether charge ($n-2$)-form ${\bf Q}[V]$ such that
%
\begin{equation}
 {\bf j}[V] = d{\bf Q}[V]. \label{eqn:def-Q}
\end{equation}


Hereafter, we assume that in an asymptotically flat spacetime there exists 
an ($n-1$)-form ${\bf B}$ such that 
%
\begin{equation}
 \int_{\infty}V\cdot\delta{\bf B}(\phi) = 
	\int_{\infty}V\cdot{\bf\Theta}(\phi,\delta\phi).
\end{equation}
By using ${\bf B}$, we can write a Hamiltonian $H[V]$ corresponding
to evolution by $V^a$ as follows~\cite{Wald1993}.
%
\begin{equation}
 H[V] \equiv \int_{\infty}({\bf Q}[V]-V\cdot{\bf B}).
\end{equation}


The symplectic current density 
${\bf\omega}(\phi,\delta_1\phi,\delta_2\phi)$ is defined by 
%
\begin{equation}
 {\bf\omega}(\phi,\delta_1\phi,\delta_2\phi) \equiv 
 	\delta_1[{\bf\Theta}(\phi,\delta_2\phi)] -
 	\delta_2[{\bf\Theta}(\phi,\delta_1\phi)]
\end{equation}
and is linear both in $\delta_1\phi$ and its derivatives, and
$\delta_2\phi$ and its derivatives~\cite{Lee&Wald1990}.


Now we define a space of solutions in which we take a variation to derive 
the first law.

Let $\tilde{\xi}^a$ be a fixed vector field, which vanishes on a
($n-2$)-surface $\Sigma$.
(Note that $\tilde{\xi}^a$ and $\Sigma$ can be defined without
referring to any dynamical fields, eg. the metric $g_{ab}$.)
In the following arguments, we consider a space of stationary, 
asymptotically flat solutions of the equations of motion 
${\bf E}(\phi)=0$, 
each of which satisfies the following three conditions.
(a) There exists a bifurcating Killing horizon with the bifurcation
surface $\Sigma$.
(b) $\tilde{\xi}^a$ is a generator Killing field of the Killing horizon. 
(c) Surface gravity corresponding to $\tilde{\xi}^a$ is $1$:
%
\begin{equation}
 \tilde{\xi}^b\nabla_b\tilde{\xi}^a = \tilde{\xi}^a,
\end{equation}
on the Killing horizon.

For each element in this space, there exist constants $\kappa$ and
$\Omega_H^{(\mu)}$ ($\mu=1,2,\cdots$) such that 
%
\begin{equation}
 \kappa\tilde{\xi}^a = t^a + \Omega_H^{(\mu)}\varphi_{(\mu)}^a,
	\label{eqn:xi-t-varphi}
\end{equation}
where $t^a$ is the timelike Killing field of stationarity with 
unit norm at infinity, $\{\varphi_{(\mu)}^a\}$ ($\mu=1,2,\cdots$) is 
a family of axial Killing fields. Hence, $\kappa$
is surface gravity and $\Omega_H^{(\mu)}$ are angular velocities of 
the horizon.

Note that, by definition, the vector field $\tilde{\xi}^a$ is fixed
under a variation of dynamical fields. We express this explicitly by
denoting the variation by $\tilde{\delta}$:
%
\begin{equation}
 \tilde{\delta}\tilde{\xi}^a = 0. \label{eqn:deltaxi=0}
\end{equation}


We now derive the first law of black hole mechanics.

First, by taking a variation of the definition (\ref{eqn:j-def}) for 
${\bf j}[\tilde{\xi}]$ and using (\ref{eqn:deltaxi=0}) and
(\ref{eqn:Theta-def}), we obtain
%
\begin{eqnarray}
 \tilde{\delta} {\bf j}[\tilde{\xi}] & = &
 	\tilde{\delta}\left({\bf\Theta}
	(\phi,{\cal L}_{\tilde{\xi}}\phi)\right)
 	- \tilde{\xi}\cdot\left({\bf E}(\phi)\tilde{\delta}\phi 
	+ d{\bf\Theta}(\phi,\tilde{\delta}\phi)\right)
 	\nonumber\\
 	& = & 
 	{\bf\omega}
	(\phi,\tilde{\delta}\phi,{\cal L}_{\tilde{\xi}}\phi)
 	+ d\left(\tilde{\xi}\cdot{\bf\Theta}
	(\phi,\tilde{\delta}\phi)\right). 
	\label{eqn:delta-j}
\end{eqnarray}
Here we have used the equations of motion ${\bf E}(\phi)=0$ and the 
following identity for an arbitrary vector $V^a$ and an arbitrary
differential form ${\bf\Lambda}$ to obtain the last line.
%
\begin{equation}
 {\cal L}_V{\bf\Lambda} = V\cdot d{\bf\Lambda} + d(V\cdot{\bf\Lambda}).
 \label{eqn:identity}
\end{equation}
Since 
${\bf\omega}(\phi,\tilde{\delta}\phi,{\cal L}_{\tilde{\xi}}\phi)$ is
linear in ${\cal L}_{\tilde{\xi}}\phi$ and its derivatives, we obtain 
%
\begin{equation}
 d(\tilde{\delta}{\bf Q}[\tilde{\xi}]) = 
 	d\left(\tilde{\xi}\cdot
	{\bf\Theta}(\phi,\tilde{\delta}\phi)\right)
 	\label{eqn:d-delta-Q}
\end{equation}
by using ${\cal L}_{\tilde{\xi}}\phi=0$ and Eq.~(\ref{eqn:def-Q}).

Next we ingrate Eq.~(\ref{eqn:d-delta-Q}) over an asymptotically flat 
spacelike hypersurface ${\cal C}$, which is orthogonal to $t^a$ at
infinity and the interior boundary of which is $\Sigma$. Since
$\tilde{\xi}^a=0$ on $\Sigma$, we obtain 
%
\begin{equation}
 \tilde{\delta}\int_{\Sigma}{\bf Q}[\tilde{\xi}] = 
	\tilde{\delta}H[\tilde{\xi}].
	\label{eqn:pre-1st-law}
\end{equation}

Finally we transform the r.h.s. and the l.h.s. of
(\ref{eqn:pre-1st-law}) to forms useful to be estimated at infinity
and the horizon, respectively.

A relation among variations of $\kappa$, $\Omega_H^{(\mu)}$, $t^a$
and $\varphi^a_{(\mu)}$ is obtained by substituting
(\ref{eqn:xi-t-varphi}) to (\ref{eqn:deltaxi=0}). 
%
\begin{equation}
 t^a\tilde{\delta}\left(\frac{1}{\kappa}\right) 
	+ \varphi^a_{(\mu)}\tilde{\delta}
	\left(\frac{\Omega_H^{(\mu)}}{\kappa}\right) =
 - \frac{1}{\kappa}\tilde{\delta}t^a
	- \frac{\Omega_H^{(\mu)}}{\kappa}
	\tilde{\delta}\varphi^a_{(\mu)}.
	\label{eqn:delta-Omega-t-varphi}
\end{equation}
By using this relation and the fact that  $H[V]$ is linear in the
vector field $V$, we can rewrite the r.h.s. of (\ref{eqn:pre-1st-law}) 
as follows.   
%
\begin{eqnarray}
 \tilde{\delta}  H[\tilde{\xi}] & = &
 	\frac{1}{\kappa}(\tilde{\delta} H[t] -H[\tilde{\delta} t]) +
 	\frac{\Omega_H^{(\mu)}}{\kappa}
	(\tilde{\delta} H[\varphi_{(\mu)}] -
 	H[\tilde{\delta} \varphi_{(\mu)}]) \nonumber\\
 & = &  \frac{1}{\kappa}\delta_{\infty}H[t] +
 	\frac{\Omega_H^{(\mu)}}{\kappa}
	\delta_{\infty}H[\varphi_{(\mu)}],
\end{eqnarray}
where the variation $\delta_{\infty}$ is defined for linear 
functionals $F[t]$ and $G_{(\mu)}[\varphi_{(\mu)}]$ so that 
%
\begin{eqnarray}
 \delta_{\infty}F[t] & = & 
 	\tilde{\delta} F[t] -F[\tilde{\delta} t],\nonumber\\
 \delta_{\infty}G_{(\mu)}[\varphi_{(\mu)}] & = & 
	\tilde{\delta} G_{(\mu)}[\varphi_{(\mu)}] -
	G_{(\mu)}[\tilde{\delta} \varphi_{(\mu)}].
	\label{eqn:def-delta-infty}
\end{eqnarray}
This newly introduced variation corresponds to a variation at 
infinity such that $t^a$ and $\varphi^a$ are fixed:
%
\begin{equation}
 \delta_{\infty}t^a =\delta_{\infty}\varphi^a_{(\mu)} = 0.
\end{equation}

In Ref.~\cite{Iyer&Wald1994} a useful expression of the Noether 
charge was given as follows.
%
\begin{equation}
 {\bf Q}[V] = {\bf W}_c(\phi)V^c + 
	{\bf X}^{cd}(\phi)\nabla_{[c}V_{d]}
	+{\bf Y}(\phi,{\cal L}_V\phi) + d{\bf Z}(\phi,V),
 	\label{eqn:Noether}
\end{equation}
where ${\bf W}_c$, ${\bf X}^{cd}$, ${\bf Y}$ and ${\bf Z}$ are locally 
constructed covariant quantities. In particular, 
${\bf Y}(\phi,{\cal L}_V\phi)$ is linear in 
${\cal L}_V\phi$ and its derivatives, and ${\bf X}^{cd}$ is given by 
%
\begin{equation}
 \left({\bf X}^{cd}(\phi)\right)_{c_3\cdots c_n} = 
 	-E_R^{abcd}{\bf\epsilon}_{abc_3\cdots c_n}.
\end{equation}
Here $E_R^{abcd}$ is the would-be equations of motion 
form~\cite{Iyer&Wald1994} for the Riemann tensor $R_{abcd}$ and 
${\bf\epsilon}_{abc_3\cdots c_n}$ is the volume $n$-form.

By using the form of ${\bf Q}$ we can rewrite the integral in the
l.h.s. of (\ref{eqn:pre-1st-law}) as 
%
\begin{equation}
 \int_{\Sigma}{\bf Q}[\tilde{\xi}] = 
	\int_{\Sigma}{\bf X}^{cd}(\phi)\nabla_{[c}\tilde{\xi}_{d]},
	\label{eqn:intQ}
\end{equation}
where we have used the Killing equation 
${\cal L}_{\tilde{\xi}}\phi=0$ and the fact that $\tilde{\xi}^a=0$ on
$\Sigma$.

Using the relation
%
\begin{equation}
 \nabla_c\tilde{\xi}_d = {\bf\epsilon}_{cd}
\end{equation}
on $\Sigma$, for any stationary solutions we can eliminate explicit
dependence of Eq.~(\ref{eqn:intQ}) on $\tilde{\xi}$, 
where ${\bf\epsilon}_{cd}$ is the binormal to 
$\Sigma$. Hence, at least within the space of stationary solutions, we
can take the variation $\tilde{\delta}$ of the integral without any
difficulties.

Therefore, we obtain the first law for stationary black holes by
rewriting Eq.~(\ref{eqn:pre-1st-law}) as
%
\begin{equation}
 \frac{\kappa}{2\pi}\tilde{\delta}S = 
	\delta_{\infty}{\cal E} - 
	\Omega_H^{(\mu)}\delta_{\infty}{\cal J}_{(\mu)},
	\label{eqn:1st-law}
\end{equation}
where entropy $S$ is defined by 
%
\begin{equation}
 S \equiv 2\pi\int_{\Sigma}{\bf X}^{cd}(\phi){\bf\epsilon}_{cd},
 	\label{eqn:S1}
\end{equation}
and energy ${\cal E}$ and angular momenta ${\cal J}_{(\mu)}$ are
defined by 
%
\begin{eqnarray}
 {\cal E} & \equiv & H[t] = 
 	\int_{\infty}({\bf Q}[t]-t\cdot{\bf B}),\nonumber\\
 {\cal J}_{(\mu)} & \equiv & -H[\varphi_{(\mu)}] = 
 	-\int_{\infty}{\bf Q}[\varphi_{(\mu)}].
	\label{eqn:def-E-J}
\end{eqnarray}

Note that, in the r.h.s. of Eq.~(\ref{eqn:1st-law}), variations of 
${\cal E}$ and ${\cal J}_{(\mu)}$ are taken with $t^a$ and 
$\varphi_{(\mu)}^a$ are fixed. This condition is explicitly implemented 
by the definition (\ref{eqn:def-delta-infty}) of $\delta_{\infty}$ 
and is necessary for a proper interpretation of the first law.


We conclude this section by giving another expression of the 
entropy.

Since $\tilde{\xi}^a$ is a generator Killing field of the Killing 
horizon, we have ${\cal L}_{\tilde{\xi}}\phi=0$ and the pull-back of 
$\tilde{\xi}\cdot{\bf L}(\phi)$ to the horizon vanishes. 
Hence, the definition (\ref{eqn:j-def}) says that the pull-back of 
${\bf j}[\tilde{\xi}]$ to the horizon is zero~\cite{JKM1994}.   
Thus, the integral of ${\bf Q}[\tilde{\xi}]$ is independent of the 
choice of the horizon cross section.

Moreover, it can be shown that the integral in (\ref{eqn:S1}) is 
the same even if we replace the integration surface $\Sigma$ by an
{\it arbitrary} horizon cross section
$\Sigma'$~\cite{JKM1994}. Therefore we obtain 
%
\begin{equation}
 S = 2\pi\int_{\Sigma'}{\bf X}^{cd}(\phi){\bf\epsilon}'_{cd},
\end{equation}
where ${\bf\epsilon}'_{cd}$ denotes the binormal to $\Sigma'$.

\section{Non-stationary perturbation}
	\label{sec:non-stationary}
	
In this section, we shall derive the first law for a non-stationary 
perturbation about a stationary black hole with a bifurcating
Killing horizon. 
Unfortunately, for non-stationary perturbations, $\delta\kappa$ and 
$\delta\Omega_H^{(\mu)}$ do not have meaning of perturbations of
surface gravity and angular velocity of the Killing horizon, even if
they are defined. 
However, since the first law (\ref{eqn:1st-law}) does not refer to
$\delta\kappa$ and $\delta\Omega_H^{(\mu)}$ but only to the
unperturbed values of $\kappa$ and $\Omega_H^{(\mu)}$, we expect that 
the first law holds also for non-stationary perturbations. 
In the following, we shall show that it does hold.


First, we specify a space of solutions in which we take a variation.

Let $\tilde{\xi}_0^a$ be a fixed vector field, which vanishes on an
fixed ($n-2$)-surface $\Sigma$.
In this section, we consider a space of asymptotically flat solutions of
the field equation ${\bf E}(\phi)=0$, for each of which
$\tilde{\xi}_0^a$ is an asymptotic Killing field.

For each solution in this space, there exist constants $\kappa$ and
$\Omega_H^{(\mu)}$ ($\mu=1,2,\cdots$) such that at spatial infinity 
%
\begin{equation}
 \kappa\tilde{\xi}_0^a = t^a + \Omega_H^{(\mu)}\varphi_{(\mu)}^a,
\end{equation}
where $t^a$ is a timelike asymptotic Killing field with unit norm at 
infinity, $\{\varphi_{(\mu)}^a\}$ ($\mu=1,2,\cdots$) is a family of
axial asymptotic Killing fields orthogonal to $t^a$ at infinity and 
$\{\Omega_H^{(\mu)}\}$ is a family of constants. 
Note that the constants $\kappa$ and $\Omega_H^{(\mu)}$ do not have 
meaning of surface gravity and angular velocities unless we consider a
stationary solution.
Moreover, in general, $\tilde{\xi}_0^a$ and $\Sigma$ have no meaning but 
an asymptotic Killing field and a fixed ($n-2$)-surface, respectively.

Note that, by definition, the vector field $\tilde{\xi}_0^a$ is fixed
under the variation. We denote the variation by $\tilde{\delta}$: 
%
\begin{equation}
 \tilde{\delta}\tilde{\xi}_0^a = 0.
\end{equation}
On the contrary, $t^a$, $\varphi_{(\mu)}^a$, $\kappa$ and
$\Omega_H^{(\mu)}$ are not fixed under the variation since definitions
of them refer to dynamical fields, which are varied. Their variations
are related by (\ref{eqn:delta-Omega-t-varphi}).

Suppose that an element $\phi_0$ of the space of solutions
satisfies the following three conditions.
(a') $\phi_0$ is a stationary solution with a bifurcating Killing
horizon with the bifurcation surface $\Sigma$. 
(b') $\tilde{\xi}_0^a$ is a generator Killing field of the Killing
horizon of $\phi_0$. 
(c') Surface gravity of $\phi_0$ corresponding to $\tilde{\xi}_0^a$ is
$1$: 
%
\begin{equation}
 \tilde{\xi}_0^b\nabla_b\tilde{\xi}_0^a = \tilde{\xi}_0^a,
\end{equation}
on the Killing horizon.


Now we derive the first law for the {\it non-stationary} perturbation 
$\tilde{\delta}\phi$ around the stationary solution $\phi_0$.

First, we mention that the validity of Eq.~(\ref{eqn:pre-1st-law}) in
the previous section depends on the following three facts.
(i) The equations of motion ${\bf E}(\phi)=0$ hold for both
unperturbed and perturbed fields. 
(Unless they hold also for perturbed fields, $\tilde{\delta}{\bf j}$
can not be rewritten as $d(\tilde{\delta}{\bf Q})$.)
(ii) $\tilde{\xi}^a$ (corresponding to $\tilde{\xi}_0^a$) is a Killing 
field of the unperturbed solution.
(iii) $\tilde{\xi}^a=0$ (corresponding to $\tilde{\xi}_0^a=0$) on 
$\Sigma$ for unperturbed solution.

These three are satisfied for the unperturbed solution $\phi_0$ and
the non-stationary variation $\tilde{\delta}\phi$ around 
$\phi_0$, too. Thus, Eq.~(\ref{eqn:pre-1st-law}) is
valid.

Since $\tilde{\delta}t^a$, $\tilde{\delta}\varphi^a_{(\mu)}$,
$\tilde{\delta}\kappa$  and $\tilde{\delta}\Omega_H^{(\mu)}$ are
related by  Eq.~(\ref{eqn:delta-Omega-t-varphi}), we can transform the
r.h.s. of (\ref{eqn:pre-1st-law}) to obtain
%
\begin{equation}
 \kappa\tilde{\delta}\int_{\Sigma}{\bf Q}[\tilde{\xi}_0] =
	\delta_{\infty}{\cal E} - 
	\Omega_H^{(\mu)}\delta_{\infty}{\cal J}_{(\mu)},
	\label{eqn:pre-1st-law2}
\end{equation}
where, as in the previous section, energy ${\cal E}$ and angular
momenta ${\cal J}_{(\mu)}$ are defined by (\ref{eqn:def-E-J}), and 
the variation $\delta_{\infty}$ is defined at infinity so that $t^a$
and $\varphi_{(\mu)}^a$ are fixed. 
Here note that $\kappa$ and $\Omega_H^{(\mu)}$ are surface gravity and 
angular velocities, respectively, for $\phi_0$.

Up to this point we have not yet used explicitly the fact that 
$\tilde{\xi}_0^a=0$ on $\Sigma$ for the perturbed solution, 
although we have used it implicitly. 
By using it explicitly, we can rewrite the l.h.s. of
(\ref{eqn:pre-1st-law2}) in a useful form. The result is 
%
\begin{equation}
 \tilde{\delta}\int_{\Sigma}{\bf Q}[\tilde{\xi}_0] =
	\frac{1}{2\pi}\tilde{\delta}S,
\end{equation}
where $S$ is defined by (\ref{eqn:S1}). 
(For explicit manipulations, see the proof of Theorem 6.1 of
Iyer-Wald~\cite{Iyer&Wald1994}.)

Finally, we obtain the first law (\ref{eqn:1st-law}) for 
non-stationary perturbations $\tilde{\delta}\phi$ about a stationary 
black hole solution $\phi_0$.


Now we comment on entropy for the perturbed, non-stationary black 
hole.

As stated above, the ($n-2$)-surface $\Sigma$ has no meaning for the 
perturbed solution. (It is a surface on which $\tilde{\xi}_0^a$ vanishes.)
In general, it does not lie on the event (or 
apparent) horizon for the perturbed solution. Hence, entropy 
evaluated on $\Sigma$ may not coincide with that on a cross 
section of the perturbed horizon, provided that 
the entropy is defined as $2\pi$ times an integral of 
${\bf Q}[\tilde{\xi}_0]$ for both $(n-2)$-surfaces. 
Note that it is in general impossible to make gauge transformation so 
that $\Sigma$ lie on a horizon cross section, if entropy 
(eg. a quarter of area in general relativity) on $\Sigma$ is 
different from entropy on a horizon cross section. 
The difference is given by $2\pi$ times an integral of the Noether 
current ${\bf j}[\tilde{\xi}_0]$ over a hypersurface whose boundary is a 
union of $\Sigma$ and a cross section of the perturbed horizon. 
Since it is natural to assign black hole entropy to the horizon 
cross section~\cite{Iyer&Wald1994}, it might be expected that there 
appears an extra term corresponding to the integral of 
${\bf j}[\tilde{\xi}_0]$ in the first law.

However, as shown in the next paragraph, the integral of 
${\bf j}[\tilde{\xi}_0]$ vanishes to first order in 
$\tilde{\delta}\phi$~\cite{Wald}.
Thus, $\tilde{\delta}S$ evaluated on $\Sigma$ gives the correct 
variation of entropy defined on the horizon to first order in
$\tilde{\delta}\phi$. 
This means that the extra term does not appear and that the first law
of Ref.~\cite{Iyer&Wald1994} derived in this section for
non-stationary perturbation about a stationary black hole 
is the correct formula.

Let us show the above statement. 
Since $\tilde{\xi}_0^a=0$ on $\Sigma$ and 
${\cal L}_{\tilde{\xi}_0}\phi_0=0$, the Noether current 
${\bf j}[\tilde{\xi}_0]$ vanishes on $\Sigma$ for the unperturbed 
solution by the definition~(\ref{eqn:j-def}). Hence, for the perturbed 
solution, the Noether current is at least first order in
$\tilde{\delta}\phi$ on $\Sigma$. On the other hand, deviation of a 
horizon cross section from $\Sigma$ is at least first order. 
Therefore, the integral of ${\bf j}[\tilde{\xi}_0]$ over a 
hypersurface connecting $\Sigma$ and the perturbed horizon cross
section is at least second order in $\tilde{\delta}\phi$.

Finally, let us apply the first law of this section to a stationary 
perturbation. The result is the same as that derived in the previous 
section. 
It is evident that the gauge condition used in this section is weaker 
than that used in the previous section. In fact, $\tilde{\xi}^a$ 
($\ne\tilde{\xi}_0^a$ for a perturbed solution) is
not fixed in the former condition. 
Hence, it can be concluded that the minimal set of gauge conditions 
necessary for the derivation of the first law is that $t^a$ and 
$\varphi_{(\mu)}^a$ are fixed at spatial infinity.

\section{Summary and discussion}
	\label{sec:summary}

In this paper we have re-analyzed Wald and Iyer-Wald derivation of 
the first law of black hole mechanics. 
In particular, two issues listed in Sec.~\ref{sec:intro} have been 
discussed in detail: 
(a) gauge conditions and (b) near-stationary black hole entropy. 
We can conclude that there are no difficulties in the derivation of the
Noether charge form of the first law for both stationary and 
non-stationary perturbations about a stationary black hole.

Unfortunately, the first law investigated in this paper cannot be 
applied to a purely dynamical situation. 
However, at least in general relativity, it seems possible 
to formulate a dynamical version of the first law as a 
law of dynamics of a trapping (or apparent) 
horizon~\cite{Hayward,HMA,Mukohyama&Hayward}.

\vskip 1cm

\centerline{\bf Acknowledgments}

The author thanks Professors H. Kodama and W. Israel for their
continuing encouragement. 
He also thanks S. A. Hayward and M. C. Ashworth for helpful
discussions. 
He was supported by JSPS program for Young Scientists,
and this work was supported partially by the Grant-in-Aid for
Scientific Research Fund (No. 9809228).


\end{document}